\documentclass[aps, prc, reprint, amsmath, groupedaddress, nofootinbib]{revtex4-1}
\usepackage[utf8]{inputenc}
\usepackage{hyperref}
\usepackage{amsmath}
\usepackage{amssymb}
\usepackage{amsfonts}
\usepackage{tabularx}
\usepackage{booktabs}
\usepackage{graphicx}
\usepackage{color}
\usepackage{multirow}
\usepackage{verbatim}
\usepackage[inline]{enumitem}
\graphicspath{{fig/}}
\definecolor{theblue}{RGB}{0,50,230}
\usepackage{appendix}
\hypersetup{
  colorlinks=true,
  linkcolor=theblue,
  citecolor=theblue,
  urlcolor=theblue
} 
\usepackage[most]{tcolorbox}

\begin{abstract}
Hard probes produced in perturbative processes are excellent probes for the study of the hot and dense QCD matter created in relativistic heavy-ion collisions.
Transport theory, allowing for coupling to an evolving medium with fluctuating initial conditions, has become a powerful tool in this endeavor. 
However, the implementation of the Landau-Pomeranchuk-Migdal (LPM) effect for medium-induced parton bremsstrahlung and pair production, poses a challenge to semi-classical transport models based on Boltzmann-type transport equations.  
In this work, we investigate a possible solution to approximate the LPM effect in a ``modified Boltzmann transport" approach, including a prescription for the running coupling constant.
By fixing a numerical parameter, this approach quantitatively reproduces the rates of medium-induced parton splitting predicted by the next-to-lead-log solution of the AMY equation which is valid in the deep-LPM regime of an infinite medium.
We also find qualitative agreement of our implementation with calculations in a finite and expanding medium, but future improvements are needed for added precision at small path length.
This work benefits transport model-based studies and the usage of these models in the phenomenological extraction of the jet transport coefficient.

\end{abstract}

\begin{document}
\title{A modified-Boltzmann approach for modeling the hot QCD medium-induce splitting vertices in the deep LPM region}
\author{Weiyao Ke}
\author{Yingru Xu}
\author{Steffen A.\ Bass}
\affiliation{Department of Physics, Duke University, Durham, NC 27708-0305}
\date{\today}
\maketitle 

\section{Introduction}
The study of hard probes in relativistic heavy-ion collisions is moving towards the precision era thanks to upcoming experimental upgrades \cite{ATLAS-Collaboration:2012iwa,Abelevetal:2014dna,STAR:upgrade-hf,Adare:2015kwa,CMS:2017dec} as well as theoretical and computational advances that allow for the calculation of jet propagation in a realistic Quark-Gluon-Plasma (QGP) medium (including event-by-event fluctuating initial conditions and temperature-dependent transport coefficients) \cite{Wang:1994fx,Zakharov:1996fv,Baier:1996sk,Zakharov:1997uu,Arnold:2002zm,Gyulassy:2003mc,Kovner:2003zj,Jeon:2003gi,CasalderreySolana:2007pr,Djordjevic:2008iz,Bass:2008rv,Schenke:2009gb,Majumder:2009zu,Majumder:2010qh,Armesto:2011ht,Zapp:2011ya,Ovanesyan:2011xy,Kang:2014xsa,Cao:2016gvr,Kauder:2018cdt,Cao:2017zih}. Among the goals for this research is the characterization of the QGP medium in terms of its jet transport coefficients $\hat{q}$.

Transport models are powerful tools that hard probes can be coupled to the realistic time evolution of the medium with an event-by-event fluctuating initial condition. 
However, the numerical implementation of the QCD analog of the Landau-Pomeranchuk-Migdal (LPM) effect, poses a serious challenge to a class of widely used models based on the Boltzmann-type transport equation or those based on Langevin dynamics.
In a dense medium, multiple scatterings act coherently within the formation time ($\tau_f$) of the medium-induced parton splitting \cite{PhysRev.103.1811,Wang:1994fx,Zakharov:1996fv,Zakharov:1997uu,Baier:1996kr,Baier:1996sk}.
The resultant bremsstrahlung rate differs from those estimated assuming independent incoherent) collisions.
As a result, bremsstrahlung in a medium effectively becomes an $n$-body to $(n+1)$-body process with a finite extended time scale $\tau_f$. At high energy, $\tau_f$ can be much larger than the collisional mean-free-path $\lambda_{\textrm{el}}$, and can be comparable to the typical inverse gradient of the macroscopic quantities of the medium created in heavy-ion collisions.
These situations are particularly difficult to treat in a Boltzmann transport equation with local and few-body collision terms. 
To simplify the parton bremsstrahlung while retaining essential qualitative features, different approximations and prescriptions are used to incorporate the LPM effect into transport models \cite{Cao:2013ita,ColemanSmith:2012vr,Xu:2004mz,Zapp:2011ya,Gossiaux:2012cv,Park:thesis}.
These models are then applied to phenomenological studies and used for the extraction of physical parameters such as $\hat{q}$.
However, insufficient attention has been paid to quantitative comparisons of these approximate LPM implementations to the theoretical baselines.
Such a step is essential for the reliable extraction of jet transport properties.

In this work, we have developed an approximation scheme for the inclusion of the LPM effect into transport models, aiming for a quantitative comparison between the model and the theory.
The resultant technique is hereafter termed as ``the modified Boltzmann transport" approach.
Here we give a short overview of the approach. 
The semi-classical transport model propagates the hard partons inside a quark-gluon plasma, under the influence of both elastic collisions with the medium and incoherent parton bremsstrahlung (inelastic processes) from independent collisions.
We first focus on the simplest scenario of a hard parton propagating in an infinite medium with a constant temperature.
The LPM effect is implemented as a modification to the incoherent inelastic rates, where elastic collisions broaden the transverse momentum between the outgoing hard partons and the incoherent rate is reduced by a factor $P$.
The suppression probability $P$ is obtained from the leading-log (LL) approximation of in-medium parton splitting when the number of coherent collisions is large $N_{\textrm{coh}} = \tau_f/\lambda_{el} \gg 1$ (the deep-LPM region), and $P \propto \lambda_{el}/\tau_f$.
Moreover, a quantitative agreement with the theory can be achieved by introducing a next-to-lead-log (NLL) correction to $P$.
The modified transport model describes the rates for in-medium bremsstrahlung $q\rightarrow q+g$, $g\rightarrow g+g$, and pair production $g\rightarrow q+\bar{q}$ surprisingly well in the deep-LPM region.
Next, we apply the model to cases beyond an infinite and static medium, because the realistic medium created in heavy-ion collisions is finite, fluctuating, with a fast dropping temperature profile due to the violent expansion. 
It is true that the current approach is developed by matching to theoretical calculations in the deep LPM region assuming many coherent collisions, but for a thin medium, the role of the interference pattern between only a few coherent collisions becomes important.
In principle, one should resort to other techniques such as the opacity expansion \cite{Wiedemann:2000za,Gyulassy:1999zd} for computation in a thin medium. 
Nevertheless, we do find the current method qualitatively reproduces the theoretical calculations of path-length dependence \cite{CaronHuot:2010bp} and the medium expansion rate dependence \cite{Baier:1998yf} of the parton bremsstrahlung.

Finally, we find it instructive to compare the current method to two other Monte Carlo implementations of the medium-induced radiation processes that have been used in previous studies.
We find that subtle differences in the construction of these models can lead to an incorrect implementation of the LPM effect or introduce correlations between sequential bremsstrahlung partons that are beyond the leading-order theory used for the model.
We draw attention to the phenomenological consequences of these differences,
especially, how these differences affect the interpretation of the extracted physical parameters such as the in-medium coupling and transport coefficient $\hat{q}$.

This paper is organized as follows. 
Section \ref{section:Boltzmann} introduces the ingredients of the semi-classical transport model that is to be modified.
In section \ref{section:modified-Boltzmann}, we propose the modifications to the aforementioned model in Section \ref{section:Boltzmann} to include the LPM effect.
In section \ref{section:results} and appendix \ref{app:tune-spectrum}, we provide detailed comparisons between the simulation and the NLL solution in the deep-LPM regime.
Effects of a finite and expanding medium are investigated in section \ref{section:more}.
In section \ref{section:compare}, we compare this work to two previously used Monte-Carlo implementations of the LPM effect in transport models.
Finally, section \ref{section:summary} summarizes and discusses the future applications of the model.

\section{Semi-classical transport approach with incoherent rates}\label{section:Boltzmann}
In this section, we construct a transport model assuming independent collisions. This type of model sets the basis for our subsequent discussion regarding the inclusion of the LPM effect.

Focusing on hard partons, collisions are categorized into elastic (hard particle number conserving) and inelastic processes (hard particle number non-conserving). 
The inelastic processes are further divided into parton-splitting and parton-fusion contributions. 
The Boltzmann equation relates the evolution of the hard parton distribution function $f_H$ to these collision processes,
\begin{eqnarray}
\frac{df_H}{dt} = \mathcal{C}_{\textrm{el}}[f_H] + \mathcal{C}_{\textrm{inel}}[f_H].
\end{eqnarray}
As a remark, we have omitted denoting the dependence of the collision term on the light parton distribution functions, because we assume them to be thermal equilibrium distribution functions with Boltzmann statistics $f_0(p) = e^{-p\cdot u/T}$. 
$T$ and $u$ are the medium temperature and four-velocity that can be obtained from a medium evolution model. 
We also neglect the collisions between two or more hard partons and the back reaction from hard parton scattering to the medium due to their small population in actual high energy nuclear collisions.
As a result, the above Boltzmann equation is linearized with respect to $f_H$.

A linearized Boltzmann equation can be solved using a collision rate approach.
First, hard partons are represented by $\delta$-functions in the phase space $f_H(t,\mathbf{x},\mathbf{p}) \approx \sum \delta^{(3)}(\mathbf{p}-\mathbf{p}(t))\delta^{(3)}(\mathbf{x}-\mathbf{x}(t))$.
Each particle travels in a straight line and occasionally its momentum changes drastically due to collisions.
This way, one obtains a stochastic solution of the time evolution of $f_H$.
The collision probability per unit time (i.e., the collision rate) is 
\begin{eqnarray}
R = \frac{g_i}{2E_1}\int  \frac{d^3p_2}{2E_2(2\pi)^3} f_0(p_2)2\hat{s} \sum_i \int_{-\hat{s}}^{0}\frac{d\sigma_i}{d\hat{t}}d\hat{t}.
\label{eq:incoh_rate}
\end{eqnarray}
For simplicity, we have only shown the case that the hard parton with momentum $p_1$ collides with one medium parton with momentum $p_2$.
$\hat{s}, \hat{t}$ are the Mandelstam variables of the two-body collisions. 
$d\sigma_i/d\hat{t}$ is the differential cross-sections of channel $i$, including both elastic and inelastic contributions.

The in-medium cross-sections $d\sigma_i$ are dominated by $\hat{t}$ channel processes which diverge like $1/\hat{t}^2$ in the vacuum.
In a medium, screening effects regulate the cross-section and leave the physical quantities finite; however, the cross-section becomes complicated and its form depends on the choice of the reference frame.
Here, we borrow an elegant solution from \cite{Ghiglieri:2015ala} to separate the in-medium collisions based on the momentum transfer-$q$ between the hard parton and the medium.
The large-$q$ transfer (hard) scattering rate uses cross-sections computed in the vacuum since the medium modification to these hard modes is small.
The associated rates are obtained from equation \ref{eq:incoh_rate} while restricting $q$ to be larger than a switching scale $Q_{\textrm{cut}}$.
The small-$q$ transfer processes are frequent and soft, which allows for a diffusion approximation of its effect on the trajectory of the hard parton,
\begin{eqnarray}
\mathbf{x}(t+\Delta t) &=& \frac{\mathbf{p}}{E}\Delta t\\
\mathbf{p}(t+\Delta t) &=& \mathbf{p} - \eta_{D,S} \mathbf{p} \Delta t + \mathbf{\xi}(t) \Delta t
\end{eqnarray}
The effects of the soft-interaction are absorbed into the drag coefficients $\eta_{D,S}$ and the covariance of the thermal random force $\mathbf{\xi}$,
\begin{eqnarray}
\left\langle\xi_i(t)\xi_j(0)\right\rangle = \delta(t) \left(
\frac{p_i p_j}{p^2}\hat{q}_{L,S} + \left(
\delta_{ij}-\frac{p_i p_j}{p^2}
\right)\frac{\hat{q}_S}{2} 
\right).
\end{eqnarray}
$\hat{q}_S$ and $\hat{q}_{L,S}$ are the soft transverse and longitudinal momentum broadening coefficients, and the subscript ``$S$'' reminds us that these numbers should only contain soft contributions with $q<Q_{\textrm{cut}}$.
The switching scale $Q_{\textrm{cut}}$ is chosen to be greater than the Debye screening mass $m_D$,
\begin{eqnarray}
m_D^2 = \frac{4\pi \alpha_s}{3}\left(N_c+\frac{N_f}{2}\right) T^2.
\end{eqnarray}
One of the many advantages of this separation is the avoidance of complicated in-medium propagators, while still achieving a leading order accuracy with a reasonable choice of $Q_{\textrm{cut}}$ at weak coupling as shown in \cite{Ghiglieri:2015ala}.
Moreover, the Lorentz-invariance of the vacuum matrix-elements used in large-$q$ processes simplifies the computation in different reference frames. The frame-dependence only appears in the diffusion equation, which is easiest solved in the medium rest frame.

\paragraph{Elastic processes} The two-body matrix-elements in the vacuum that enter the large-$q$ collision rates can be found in the standard literature \cite{RevModPhys.59.465}.
In the small-$q$ diffusion sector, the transverse and longitudinal momentum diffusion coefficients $\hat{q}_S, \hat{q}_{S,L}$ in a weakly coupled theory have been calculated in \cite{Ghiglieri:2015ala} at leading order,
\begin{eqnarray}
\hat{q}_S = \int_0^{Q_{\textrm{cut}}^2} dq^2 \frac{\alpha_s m_D^2 T}{q^2 (q^2+m_D^2)},
\label{eq:qS} \\
\hat{q}_{L,S} = \int_0^{Q_{\textrm{cut}}^2} dq^2 \frac{\alpha_s m_\infty^2 T}{q^2 (q^2+m_\infty^2)}.
\label{eq:qSL}
\end{eqnarray}
$m_{\infty}$ is the asymptotic gluon thermal mass $m_{\infty}^2 = m_D^2/2$.
Finally, the drag coefficient is determined by the Einstein relation between the transport coefficients,
\begin{eqnarray}
\eta_{D,S} = \frac{\hat{q}_{L,S}}{2ET} - \frac{d\hat{q}_{L,S}}{dp^2} - \frac{\hat{q}_{L,S} - \hat{q}_S/2}{p^2}.
\end{eqnarray}

\paragraph{Inelastic processes} The inelastic collision term is also separated into a large-$q$ $2\leftrightarrow 3$ body inelastic collision rate plus an effective $1\leftrightarrow 2$ body diffusion-induced parton splitting / fusion rate.
The matrix-elements of $2\leftrightarrow 3$ body collisions are derived under the limit $\mathbf{k}^2, \mathbf{q}^2 \ll x(1-x)\sqrt{\hat{s}}$.
Here, $x$ is the light-cone energy fraction of the initial state hard parton carried by one of the final state hard parton, $\mathbf{q}$ is the transverse component of $q$ in the center-of-mass frame of the collision, and $\mathbf{k}$ is the transverse momentum between the two split hard partons in the final state. 
A list of these matrix-elements can be found in appendix \ref{app:23}.
Again, the rates are obtained from equation \ref{eq:incoh_rate} imposing $q>Q_{\textrm{cut}}$.

The diffusion-induced splitting rate uses the restriction $q<Q_{\textrm{cut}}$ in equation \ref{eq:incoh_rate}, and uses the limit $\mathbf{q} \ll \mathbf{k}$ of the $2\rightarrow 3$  matrix-elements listed in appendix \ref{app:23},
\begin{eqnarray}\nonumber
R_{1\rightarrow 2} &=& \frac{g_i}{2E_1}\int  \frac{d^3p_2f_0(p_2)}{2E_2(2\pi)^3} 2\hat{s} \int_{0}^{Q_{\textrm{cut}}^2} \mathbf{q}^2 \frac{d\sigma_{\textrm{el}}}{q^2} dq^2 
\\
&& \times \int d \mathbf{k}^2 dx \frac{\alpha_s P(x) }{2\pi (\mathbf{k}^2 + m_\infty^2)^2} \label{eq:rate12_1}
\end{eqnarray}
where $m_\infty$ is added to screen the divergence and $P(x)$ is the vacuum splitting function listed in appendix \ref{app:23}.
Now, one may notice that the first line in equation \ref{eq:rate12_1}, upon summing over all channels, simply computes the variance of transverse momentum received by the hard parton per unit time from soft interactions below the momentum cut-off $Q_{\textrm{cut}}$.
So we rewrite $R_{1\rightarrow 2}$ using the soft transport coefficients $\hat{q}_{S}$ as
\begin{eqnarray}
R_{1\rightarrow 2} &=& \hat{q}_S\int d \mathbf{k}^2 dx \frac{\alpha_s P(x) }{2\pi (\mathbf{k}^2 + m_\infty^2)^2} \label{eq:rate12_2},
\end{eqnarray}
which is our final expression for the diffusion induced inelastic collision rate for the incoherent transport equation.
Finally, the collision rates of the reverse processes $3\rightarrow 2$ and $2\rightarrow 1$ processes can be written down by the requirement of detailed balance.

\paragraph{The transport equation in the incoherent limit}
Combining all these processes, we summarize the semi-classical transport equation with independent collisions into
\begin{eqnarray}\label{eq:incoh_transport}
\frac{df}{dt} = \mathcal{D}[f] + \mathcal{C}_{1\leftrightarrow 2}[f] + \mathcal{C}_{2\leftrightarrow 2}[f] + \mathcal{C}_{2\leftrightarrow 3}[f].
\end{eqnarray}
The distribution function of the hard parton evolves under the effect of diffusion $\mathcal{D}$, large-$q$ elastic collision $\mathcal{C}_{2\leftrightarrow 2}$,  diffusion induced parton splitting / merging $\mathcal{C}_{1\leftrightarrow 2}$, and large-$q$ inelastic collisions $\mathcal{C}_{2\leftrightarrow 3}$.

\section{Modeling LPM effect by a modified transport simulation}\label{section:modified-Boltzmann}
The incoherent transport equation requires: first, the transition time scale of a process is small compared to the mean-free-path, so that multiple-collision contribution to the transition rate is negligible; and second, the transition time scale is small compared to the inverse-gradient of the system, so that the collision terms in equation \ref{eq:incoh_transport} depend only on distribution functions at a localized space-time region.
Such a semi-classical picture does work for elastic collisions at weak coupling $g\ll 1$. 
This is because a medium scattering center is statistically independent from others with distances greater than $1/m_D\sim 1/gT$ and the mean-free-path $\lambda \sim 1/g^2T$ is larger than $1/m_D$.
For an inelastic collision, consider the splitting of a hard parton ``$a$'' with energy $E$ to two hard partons ``$b$'' and ``$c$'', with ``$b$'' carrying an $x$ fraction of the original parton's energy.
From uncertainty principle, the formation time $\tau_f$ of the hard final state is obtained as the inverse of the light-cone energy difference $\delta E$ between the initial and final states,
\begin{eqnarray}
\tau_f^{-1} \sim \delta E = \frac{\mathbf{k}^2}{2x(1-x)E}.
\label{eq:tauf}
\end{eqnarray}
$\mathbf{k}$ is the transverse momentum of ``$b$'' relative to the direction-of-motion of ``$a$''.
For hard and collinear splittings, this formation time can be very large compared to $\lambda$ and multiple-collision effect becomes important and needs to be resummed into the transition rate.

In an infinite and static medium, when the number of multiple collisions $N$ is large (the deep-LPM region), theoretical calculation indicates a qualitative change to the parton radiation pattern comparing to the results obtained in the independent collision picture \cite{Baier:1996kr}.
First, transverse momentum of the splitting is broadened from multiple collisions; second, the transition rate is reduced from producing $\mathcal{O}(\alpha_s)$ radiation every collision to producing $\mathcal{O}(\alpha_s)$ radiation every formation time,
\begin{eqnarray}
\frac{dR}{d\omega} \propto \frac{\alpha_s}{\omega\lambda} \rightarrow \frac{\alpha_s}{\omega\tau_f}.
\end{eqnarray}
The average inverse formation time can be estimated using equation \ref{eq:tauf} and the condition $\left\langle \mathbf{k}^2 \right\rangle \approx \hat{q} \tau_f$,
because the transport coefficient $\hat{q} = d \left\langle \mathbf{k}^2 \right\rangle / dt$ quantifies the momentum broadening per unit time.
These two conditions lead to,
\begin{eqnarray}
\langle \tau_f^{-1} \rangle \sim \sqrt{\frac{\hat{q}}{2x(1-x)E}}.
\label{eq:tauf-sf}
\end{eqnarray}
At weak coupling $\hat{q}\propto g^4 T^3$ and $\tau_f/\lambda \sim \sqrt{2x(1-x)E/T}$.
The radiative pattern with moderately small $x$ in an infinite medium is changed to,
\begin{eqnarray}
\frac{dR}{d\omega} \propto  \frac{\alpha_s T^{1/2}}{\omega^{3/2}\lambda}
\end{eqnarray}
Therefore, a fundamental modification to the semi-classical equation is necessary once the final-state partons become hard $xE, (1-x)E > T$.

\subsection{The modification to the semi-classical evolution}
We start by investigating the leading-order calculation of medium-induced parton splitting in a ``brick'' medium of constant temperature to identify the modification we need.
Here, we quote the reorganized leading-order formula for the probability of a single medium-induced splitting in a brick medium from \cite{Zakharov:1996fv,CaronHuot:2010bp},
\begin{eqnarray}
\frac{dP^{a}_{bc}}{d\omega} &=& \int_0^\infty dt \frac{g^2}{\pi E} P_{bc}^{a(0)}(x) \int_t^\infty dt'  F(t', t),
\label{eq:full-theory}
\\
F(t', t) &=& \mathfrak{Re} \int_{{\bf q}', {\bf q}} \frac{i {\bf q}'\cdot {\bf q}}{\delta E} \mathcal{C}(t') \circ K(t', {\bf q}'; t, {\bf q}).
\end{eqnarray}
$P_{bc}^{a(0)}$ is the vacuum splitting function and $\delta E$ is the light-cone energy difference between initial and final states. 
The $\mathcal{C}(t')$ operator is a Boltzmann-type collision operator in the momentum space such that,
\begin{eqnarray}
\mathcal{C}\circ f_{\mathbf{p}} = \int_{\bf q} g^2\mathcal{A}(\mathbf{q}^2)
&&\left\{  \frac{C_b+C_c-C_a}{2}\left(f_{\bf p}-f_{{\bf p}-{\bf q}}\right) \right.\\\nonumber
 +&&    \frac{C_a+C_c-C_b}{2}\left(f_{\bf p}-f_{{\bf p}+x{\bf q}}\right) \\\nonumber
+&&  \left. \frac{C_a+C_b-C_c}{2}\left(f_{\bf p}-f_{{\bf p}+(1-x){\bf q}}\right)\right\},
\end{eqnarray}
where $\omega$ is energy of daugther parton ``$b$''.
$C_i$ is the color factor of each parton and $\int_{\mathbf{q}}$ represents an integration over transverse momentum $\int d\mathbf{q}^2/(2\pi)^2$.
The collision kernel is given by \cite{Aurenche:2002pd},
\begin{eqnarray}
\mathcal{A}(\mathbf{q}^2) = \frac{m_D^2 T}{q^2\left(m_D^2+q^2\right)}.
\label{eq:kernel}
\end{eqnarray}
Finally, $K(t', {\bf q}'; t, {\bf q})$ is the propagator of the transverse  Hamiltonian $\hat{H} = \delta E - i\mathcal{C}$ in the momentum representation.
This rather compact result is actually hard to implement in a Boltzmann formulation, because of the double-time integral that comes from the nature of a quantum transition.
Moreover, the splitting probability generally dependents on the temperature and flow velocity profiles of the medium, making it a computationally heavy task when coupled to dynamical evolving and fluctuating medium.

Our approximation towards a modified Boltzmann transport formulation starts by replacing the effect of the temporal two-point function $F(t',t)$ with a simple ansatz,
\begin{eqnarray}
F(t', t) \rightarrow \frac{1}{N}\sum_{i=1}^N \frac{b}{\tau_i(t)} \delta(t-t'- a \tau_i(t)).
\label{eq:ansatz}
\end{eqnarray}
Here the function $F(t', t)$ is approximated by an ensemble of $N$ independent copies of the system $a\rightarrow b+c$.
These copies are generated according to the incoherent $1\rightarrow 2$ and $2\rightarrow 3$ rates as hard parton $a$ propagates.
Each copy ``$i$" evolves with the influence of the elastic broadening. 
Its formation time $\tau_i(t)$ at time $t$ during the evolution can be computed by equation \ref{eq:tauf}.
The delta function imposes that the branching that starts at time $t'$  is thought to be formed at time $t+a\tau_f$.
The additional factor $b/\tau_f$ accounts for that the branching probability is suppressed compared to the incoherent case.
This ansatz of representing the function $F(t', t)$ with information at $t$ and $t'$ of an ensemble of particles follows the same spirit of representing the distribution function by an ensemble of particle states.
Of course, this is only a crude ansatz for $F(t', t)$, as the latter is actually highly oscillating, and the validity has to be examined by comparing its prediction with the theoretical calculations.
Finally, $a$ and $b$ are dimensionless factors whose forms shall be determined in later comparison with theory and will be tuned to achieve an optimal level of agreement to theoretical calculations.

With such an ansatz, the medium-induced splitting probability reduces to,
\begin{eqnarray}
\nonumber
\frac{dP^{a}_{bc}}{d\omega} &=& \int_0^\infty dt \frac{g^2 P_{bc}^{a(0)}}{\pi E\tilde{\lambda}(t)} \frac{1}{N}\sum_{i=1}^N \int_t^\infty dt' \frac{b \tilde{\lambda}(t)}{\tau_i(t)} \delta(t-t'- a \tau_i(t)) \\
&=& \frac{1}{N}\sum_{i=1}^N\int_0^\infty dt \frac{dR_{\textrm{incoh}}(t)}{d\omega} \times \left.\frac{ab\tilde{\lambda}(t)}{\tau_f(t',t)}\right|_{t'=t+a\tau_f}
\label{eq:procedure}
\end{eqnarray}
where we have divided and multiplied back an effective mean-free-path $\tilde{\lambda}(t) = m_D^2/\hat{q}_g$ so that we may interpret the quantity immediately after the $dt$ integral as the incoherent splitting rate $R_{\textrm{incoh}}$.
The reason for using an effective $\tilde{\lambda}$ is that it can be defined for both scattering and diffusion processes, provided only the screening mass and the transverse momentum diffusion coefficient.
From equation \ref{eq:procedure}, it is clear how we can modify the standard Boltzmann transport to include the LPM effect in the deep-LPM region approximately.
During the simulation of the incoherent transport  (equation \ref{eq:incoh_transport}) of each hard parton, we implement the following modification the parton bremsstrahlung and pair production processes.
Suppose a parton splitting process ``$a\rightarrow b+c$'' happens at $t=t'$,
\begin{enumerate}
\item Final state partons $b$ and $c$ are not immediately treated as physical objects (``preformed'') to the system. A parton $a$ can carry arbitrary numbers of such ``preformed'' final-state copies.
\item Evolve from $t$ to $t+\Delta t$ the ``preformed'' final state partons in each copy with only elastic processes. 
While the mother parton $a$ is evolved under the full collision term in equation \ref{eq:incoh_transport}.
\item Recalculate formation time $\tau_f$ after each time step. 
The elastic broadening from multiple collisions in step 2, on average, shortens the formation time.
\item Repeat steps 1 to 3 until $t-t' > a\tau_f(t)$ is satisfied. 
\item Then, the ``performed'' final state is considered to become the physical final state with probability $p$, 
\begin{eqnarray}
p = \min\left\{1, \frac{ab\tilde{\lambda}(t)}{\tau_f}\right\}
\label{eq:rejection}
\end{eqnarray}.
\end{enumerate}
In the last step, daughter partons from accepted final state will be treated as independent objects thereafter and be propagated by the full transport equation.
Rejected final states are dropped and do not cause any physical effect. 
A key step is a self-consistent determination of the formation time as described in step 4. 
In a static medium, it results in the expected scaling of the average inverse formation time in equation \ref{eq:tauf-sf}.
This procedure also generalizes to medium with evolving temperature and flow velocity profiles, as the elastic broadening is performed along the trajectory of the hard parton.
This iterative procedure was first developed and implemented by \cite{Zapp:2011ya}.

In the following subsection, we shall compare such modification to the leading-log (LL) and the next-to-leading-log (NLL) approximation of splitting rates in an infinite medium. 
We will show that this modification indeed reproduces the qualitative features given by the lead-log results once $a$ is related to the color factors, while the NLL results help to determine the form of $b$ to achieve a quantitative agreement with the theory.

\subsection{Matching the modified transport equation to theoretical calculations in an infinite medium}
Going to an infinitely large medium with an uniform temperature, equation \ref{eq:full-theory} can be cast into its asympototic form known as the AMY formalisim \cite{Arnold:2002ja,Arnold:2002zm,Arnold:2003zc},
\begin{eqnarray}\label{eq:AMY-1}
\frac{dR^a_{bc}}{d\omega} &=& \frac{\alpha_s d_a P^{a(0)}_{bc}(x)}{E\nu_a} \int\frac{d^2\mathbf{k}}{(2\pi)^2} \frac{2\mathbf{k}\cdot \mathfrak{Re} \mathbf{F}}{4x^2(1-x)^2}
\end{eqnarray}
where we have dropped the Bose enhancement and the Pauli blocking factors from the original formula.
The vector-valued function $\mathbf{F}(\mathbf{k}; E, x)$ satisfies the following static equation,
\begin{eqnarray}\label{eq:AMY-2}
2\mathbf{k} &=& i\frac{\mathbf{k}^2 \mathbf{F}(\mathbf{k})}{2x(1-x)E} + g^2 \mathcal{C}[\mathbf{F}]
\end{eqnarray} 
The exact solution can be solved numerically and has already been applied to transport study \cite{Jeon:2003gi,Schenke:2009gb}, but we shall use approximated semi-analytic solutions obtained in \cite{Arnold:2008zu}.
These approximated solutions were obtained at the leading-log (LL) and next-to-leading-log (NLL)accuracy, and the NLL solution was shown to be a good approximation of the numerical results in the deep LPM regime $\omega \gg T$.

At leading-log order, a small-$q$ expansion was performed to obtain a diffusion approximation to the operator $\mathcal{C}$ below a certain cut-off $q<Q_0$.
The resulting leading-log solution is \cite{Arnold:2008zu},
\begin{eqnarray}\label{eq:AMY-LL}
\frac{dR_{bc}^{a,\textrm{LL}}}{d\omega} &=& \frac{\alpha_s P_{bc}^{a(0)}}{\pi E\sqrt{2}}
\sqrt{\frac{\hat{q}_3(x, Q_0^2)}{2x(1-x)E}}
\end{eqnarray}
Where the $\hat{q}_3$ is defined as an effective transport parameter,
\begin{eqnarray}
\hat{q}_3(x, Q_0^2) &=& C_{abc}(x) \int_{\mathbf{q}^2 < Q_0^2} g^2\mathcal{A}(\mathbf{q}^2) \mathbf{q}^2 \frac{d\mathbf{q}^2}{(2\pi)^2} \\
&=& C_{abc}(x) \alpha_s T m_D^2 \ln\left(1+\frac{Q_0^2}{m_D^2}\right).\label{eq:qhat3}
\end{eqnarray}
Note that $\hat{q}_3$ is logarithmically dependent on the cut-off $Q_0$.
It comes from integrating the perturbative tail of the collision kernel in equation \ref{eq:kernel}.
$C_{abc}(x)$ is a process- and $x$-dependent color factor,
\begin{eqnarray}
C_{abc}(x) &=&  \frac{C_b+C_c-C_a}{2} + x^2 \frac{C_a+C_c-C_b}{2} \\\nonumber
&+& (1-x)^2\frac{C_a+C_b-C_c}{2}.
\end{eqnarray}

Comparing the leading-log formula to the modified Boltzmann approach in equation \ref{eq:procedure}, the term $\sqrt{\hat{q}_3 / 2x(1-x)E}$ in equation \ref{eq:AMY-LL} plays the role of the inverse formation time.
We can also insert $1/\tilde{\lambda} \times \tilde{\lambda}$ into equation \ref{eq:AMY-LL} so that it resembles the modified rate in equation \ref{eq:procedure}.
However, the effective $\hat{q}_3$ differs from the $\hat{q}$ of a single particle, e.g. particle ``$b$'' on which we performed the elastic broadening, by an $x$ and color-dependent factor $C_{abc}$.
This can be improved by using a process- and $x$-dependent $a$ parameter in equation \ref{eq:ansatz},
\begin{eqnarray}
a \rightarrow a_{abc}(x) = \frac{C_b}{C_{abc}(x)}
\end{eqnarray}

Another issue is that the LL solution in equation \ref{eq:AMY-LL} is still ambiguous up to an unknown cut-off $Q_0$ through $\hat{q}_3(x, Q_0^2)$.
This $Q_0$ ambiguity can be improved by going to the NLL order, where the contribution of the collision kernel with $q>Q_0$ is treated as a perturbation to the LL solution.
This ``hard" correction to the ``multiple-soft" approximation is also recently studied in \cite{Mehtar-Tani:2019tvy} in the BDMPS framework.
In both works, the NLL result is expressed as the LL solution with the unknown $Q_0$ replaced by $Q_1$,
\begin{eqnarray}
Q_1^2  \approx \sqrt{\omega \hat{q}} \approx \sqrt{\omega \alpha_s C_R m_D^2 T \ln\frac{Q_0^2}{m_D^2}}
\label{eq:Q1}
\end{eqnarray}
or one can uses a self-consistent determination of $Q_1$ as in \cite{Arnold:2008zu} to eliminated the dependence of $Q_0$ from the formula,
\begin{eqnarray}
Q_1^2 &=& \sqrt{2 x (1-x) E \alpha_s T m_D^2}\\\nonumber
&\times & \left(
\frac{C_b+C_c-C_a}{2}\ln\frac{2\xi Q_1^2}{m_D^2} \right.\\\nonumber 
&+& \frac{C_a+C_c-C_b}{2} x^2 \ln\frac{2\xi Q_1^2}{x^2 m_D^2} \\\nonumber 
&+& \left.\frac{C_a+C_b-C_c}{2} (1-x)^2 \ln\frac{2\xi Q_1^2}{(1-x)^2 m_D^2} \right)^{1/2}
\label{eq:Q1-sf}
\end{eqnarray}
Therefore, the reasonable choice of the scale in $\hat{q}_3$ is of similar order as the transverse momentum of the branching $\mathbf{k}^2 \approx \sqrt{2x(1-x)E \hat{q}}$.
Now, check what this scale is in the original transport approach: the large-$Q$ two-body matrix-element in equation \ref{eq:incoh_rate} is always integrated up to the maximum transfer bounded by the center-of-mass energy $\sqrt{\hat{s}}$ of each independent collision. 
Ignore the slow $\hat{s}$-dependence of the elastic cross-section at high-energy and average over the medium parton momentum $p_2$ in $\hat{s} = (p+p_2)^2$ over the thermal distribution, then the average $Q_{0}^2$ in the transport simulation is  $\langle\hat{s}\rangle = 6ET$.
Therefore, we must correct for this difference in scale; otherwise, the ansatz with a na\"ive choice of an upper limit of the matrix-element integration leads to systematic deviation from the NLL solution in a logarithmic manner.
Noticing that the inverse formation time is proportional to $\sqrt{\ln(1+\hat{Q}^2/m_D^2)}$, a simple correction can be made using a scale-dependent $b$ parameter in the acceptance probability of the modified transport equation in equation \ref{eq:rejection},
\begin{eqnarray}
b &=& 0.75\sqrt{\frac{\ln(\hat{Q}_1^2 )}{\ln(\hat{Q}_0^2 )}},
\label{eq:NLL-b}
\end{eqnarray}
with the NLL-improved scale $\hat{Q}_1^2$ and the native choice of scale in the transport equation $\hat{Q}_0^2$ given by,
\begin{eqnarray}
\hat{Q}_1^2 &=& 1 + \frac{\mathbf{k}^2}{m_D^2} \approx 1 + \frac{\tau_{f,i}}{\tilde{\lambda}}, \\ 
\hat{Q}_0^2 &=& 1 + \frac{6ET}{m_D^2}.
\end{eqnarray}
The prefactor $0.75$ in equation \ref{eq:NLL-b} is the only numerical constant that we have tuned when we compared the modified Boltzmann simulation to the NLL solutions in the next section, and it is the same throughout the rest of the paper.
As a remark, this logarithmic correction comes from the integration of the perturbative collision kernel at large-$q^2$.
Therefore, if one assumes and implements certain collision kernel that vanishes sufficiently fast at large-$q$, this logarithm factor in $b$ should be dropped. 

\subsection{Agreement in the Bethe-Heitler region at large-$q$}
In the previous subsection, we have shown how the modified Boltzmann transport approach can be matched to the NLL solution in the deep-LPM region where $\tau_f \gg \lambda_{\textrm{el}}$.
In this subsection, we demonstrate that the transport equation agrees  with the theoretical calculation in another region of phase-space
This is the large-$q$ ($q>Q_{\textrm{cut}}$) region in the Bethe-Heitler regime $\tau_f \ll \lambda_{\textrm{el}}$.
The acceptance probability \ref{eq:rejection} goes back to unity for this region, and the transport equation is simply the original Boltzmann equation with incoherent rates.
Here, we briefly show that the incoherent $2\rightarrow 3$ rates used in the equation \ref{eq:incoh_transport} is the same as the rate $dR/d\omega$ obtained as the leading term in the $1/\tau_f$ expansion of the AMY formalism in equations \label{eq:AMY-1} and \label{eq:AMY-2}.

When the formation time is very short, one treats $1/\tau_f = \mathbf{k}^2/2x(1-x)E$ in equation \ref{eq:AMY-2} as a large number and solve for the real-part of $\mathbf{F}$ by one iteration,
\begin{eqnarray}
\int\frac{d^2\mathbf{k}}{(2\pi)^2} \frac{2\mathbf{k}\cdot \mathfrak{Re} \mathbf{F}}{4x^2(1-x)^2E^2} &=& 2g^2 \mathbf{\phi}_k\cdot \mathcal{C}[\mathbf{\phi}_k] \label{eq:BH-solution},
\end{eqnarray}
where $\mathbf{\phi}_k = \mathbf{k}/\mathbf{k}^2$.
Taking $q\rightarrow q+g$ as an example, the total $2\rightarrow 3$ rate in the Boltzmann equation is (use $\hat{s} \approx 6 ET$),
\begin{eqnarray}
\frac{dR^{q}_{qg}}{d\omega} &\propto&  \sum_i\int_{q^2 > Q_{\textrm{cut}}^2}   \frac{f_i(p_2)dp_2^3}{(2\pi)^3 2p_2}  \frac{g^4 T}{q^4} dq^2 d k^2\\\nonumber
&& \left\{
C_F\left( \mathbf{\phi}_{k-q}-\mathbf{\phi}_{k-xq} \right)^2
+ C_F\left( \mathbf{\phi}_{k-q}-\mathbf{\phi}_{k} \right)^2\right.\\\nonumber
&&\left.
- (2C_F-C_A)\left( \mathbf{\phi}_{k-q}-\mathbf{\phi}_{k-xq} \right)\cdot \left( \mathbf{\phi}_{k-q}-\mathbf{\phi}_{k} \right)
\right\}
\end{eqnarray}
where we have used the cross-section from appendix \ref{app:23} and the splitting function is not shown explicitly.
This transverse momentum integration looks different from the one in equation \ref{eq:BH-solution}.
Summing over the species, colors, and degeneracy of the particle ``$i$'', the integration of the momentum $p_2$ yields the Debye mass with classical statistics,
\begin{eqnarray}
\frac{g^4 T}{q^4}\sum_i\int \frac{f_i(p_2)dp_2^3}{(2\pi)^3 2p_2} = \frac{2g^2 m_D^2}{q^4}.
\end{eqnarray}
Expanding each squares and changing the integration variable from $k$ to one of $k-q, k-xq, k-(1-x)q$ for each term, one can show that the integration can be cast into,
\begin{eqnarray}
\frac{dR^{q}_{qg}}{d\omega} &\propto& \int_{q^2 > Q_{\textrm{cut}}^2}  \frac{2g^2 m_D^2}{q^4} d q^2 dk^2\\\nonumber
&& \left\{
C_A\mathbf{\phi}_{k}\cdot \left( \mathbf{\phi}_{k}-\mathbf{\phi}_{k+q} \right)
+C_A\mathbf{\phi}_{k} \cdot \left( \mathbf{\phi}_k - \mathbf{\phi}_{k+(1-x)q}\right) \right.\\\nonumber
&&\left.+(2C_F-C_A)\mathbf{\phi}_{k} \cdot \left(\mathbf{\phi}_k-\mathbf{\phi}_{k+xq} \right)
\right\} \\
&=&  \int dk^2 \mathbf{\phi}_k \cdot \mathcal{C}[\mathbf{\phi}_k]
\end{eqnarray}
which agrees with the large-$q^2$ limit of of the AMY solution in equation \ref{eq:BH-solution} when $q^2 \gg m_D^2$.

\subsection{Implement running of $\alpha_s$}\label{section:running}
At the end of this section, we introduce the prescription for a running coupling constant in the modified Boltzmann approach.
We used the leading order running coupling constant with $n_f = 3$ and $\Lambda = 0.2$ GeV, 
\begin{eqnarray}
\alpha_s(Q^2) = \frac{4\pi}{9\ln\left(Q^2/\Lambda^2\right)}
\end{eqnarray}
To avoid the pole when $Q$ approaches the non-perturbative scale, we introduce a cut-off at a medium scale $Q_{\textrm{med}} = \pi T$. 
Therefore the actual coupling constant is $\alpha_s(\max\{Q, Q_{\textrm{med}}\})$.
Following the prescription described in \cite{Arnold:2008zu}, the coupling constant associated to the collision kernel $\mathcal{C}$ are evaluated at $\mathbf{q}^2$ and the resulting $\hat{q}$ can be integrated approximately to get the running version of the Eq. \ref{eq:qhat3},
\begin{eqnarray}
\hat{q}_3^{\textrm{running}} \approx \frac{4\pi}{9}\left(g^2(m_D^2) - g^2(Q_0^2)\right) 1.27 T^3 C_{abc}(x)
\label{eq:q3running}
\end{eqnarray}
Where the scale $Q_0$ is of order $m_D [E/T \ln(E/T)]^{1/4}$.
And the scale of the $\alpha_s$ associated to the splitting  vertex in equation \ref{eq:AMY-LL} is chosen at the typical transverse momentum $\mathbf{k}^2 \sim \sqrt{2x(1-x)E\hat{q}_3}$.

In the modified Boltzmann approach, such a running coupling prescription is straightforward for the elastic part.
The transport coefficients $\hat{q}_S$ and $\hat{q}_{S, L}$ in Eq. \ref{eq:qS} and equation \ref{eq:qSL} for the diffusion sector should be integrated with $\alpha_s(q^2)$.
The $\alpha_s$ associated to the $2\rightarrow 2$ scattering matrix-elements (including the ones that appear in the $2\rightarrow 3$ matrix-elements) are evaluated at the $t$-channel momentum transfer squared.
The scale $\mathbf{k}^2$ for the splitting vertex coupling requires an additional treatment, because $\mathbf{k}$ comes from the summation over multiple scatterings within the formation time,
\begin{eqnarray}\label{eq:kTn}
\mathbf{k}_{t_0+\tau_f}^2 = \left(\mathbf{k}_{t_0}+\mathbf{q}_1+\cdots+\mathbf{q}_n\right)^2.
\end{eqnarray} 
However, the initial splitting processes are generated using incoherent processes with the coupling evaluated at $\mathbf{k}^2(t_0)$.
This is on average $\sqrt{\omega/T}$ times smaller than $\mathbf{k}^2(t_0+\tau_f)$ in the deep LPM region.
Therefore the effective coupling after multiple scattering is smaller than the one we used in incoherent calculation and allows for a rejection implementation by modifying the acceptance probability $p$ in equation \ref{eq:rejection} to
\begin{eqnarray}
p^{\textrm{running}} = p\times \frac{\alpha_s(k_{t_0+\tau_f}^2)}{\alpha_s(k_{t_0}^2)}.
\end{eqnarray}
This final step completes the inclusion of running coupling effect in the modified Boltzmann approach.

\section{results}\label{section:results}
\begin{figure}
\includegraphics[width=\columnwidth]{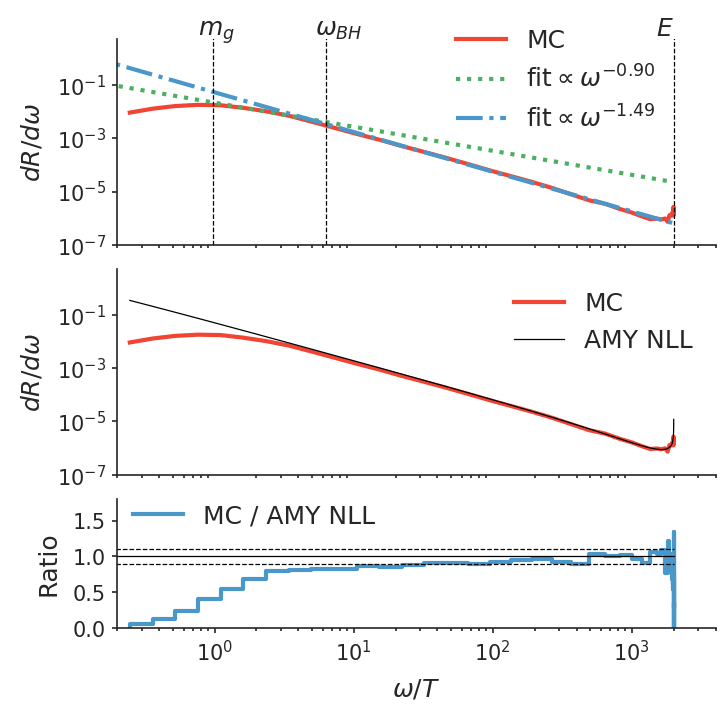}
\caption{The $q\rightarrow q+g$ splitting rate in an infinite medium from a quark with $E=1$ TeV,and a coupling constant $\alpha_s = 0.1$. The top plot shows the simulated spectrum $dR/d\omega$ (red-dashed line) and power law fit (green-dotted and blue-dash-dotted lines) in different gluon energy regions, separated by energy scales $\omega_{BH}\approx 2\pi T$. The middle plot compares to the simulation to NLL solution to the AMY equation, and the ratio is shown in the bottom plot}
\label{fig:spectrum}
\end{figure}

\begin{figure}
\includegraphics[width=\columnwidth]{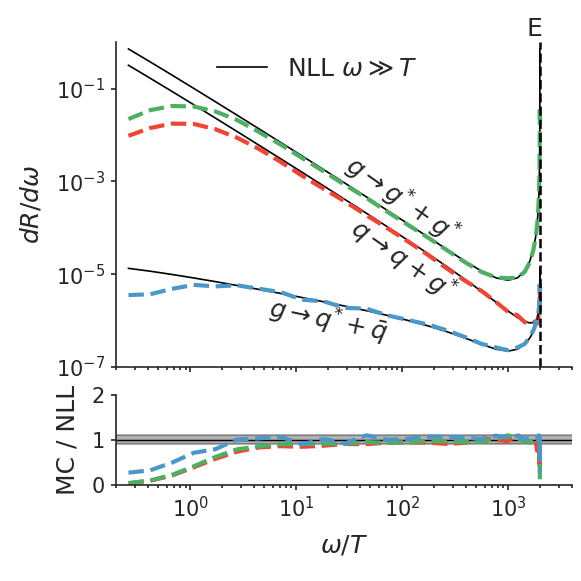}
\caption{The splitting rate of $q\rightarrow q+g^*$, $g\rightarrow g+g^*$, and $g\rightarrow q^* + \bar{q}$ as a function of the parton energy labeled by the star. The mother parton with $E=1$ TeV evolves inside an infinite medium with $T=0.5$ GeV. The simulations (thick dashed lines) are compared to the NLL solutions (thin solid lines).}
\label{fig:channel_rate}
\end{figure}

\begin{figure}
\includegraphics[width=\columnwidth]{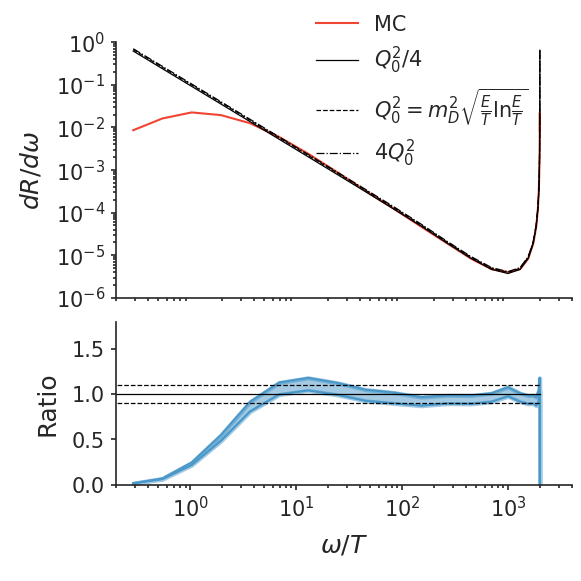}
\caption{Top plot: comparison of modified Boltzmann simulation with the NLL solution  with running coupling.
Three initial guesses of the 
The ratio between simulation and theory is shown in the bottom plot.}
\label{fig:running}
\end{figure}

In this section, we compare the rate of in-medium bremsstrahlung and pair production from the simulation of the modified Boltzmann approach to the NLL approximation of the AMY equation in an infinite medium.

The differential rate $dR/d\omega$ is shown in Figure \ref{fig:spectrum} for a 1 TeV quark splitting into a gluon and a quark.
The temperature of the medium is $T=0.5$ GeV and we choose a relatively small coupling constant $\alpha_s = 0.1$.
Please also refer to appendix \ref{app:tune-spectrum} for a full comparison varying both the parton energy and the coupling constant.
The horizontal axis is the radiated gluon energy.
In the upper plot, we divided the spectrum into different regions by the gluon thermal mass $m_\infty$ and an estimated Bethe-Heitler energy $\lambda_g m_D^2 \sim 2\pi T$.
The spectrum at $\omega < m_g$ is suppressed due to the gluon thermal mass.
In the Bethe-Heitler region $\omega < 2\pi T$, the spectrum scales like $\omega^{-1}$ as given by the incoherent radiation rate.
In the deep LPM region $T \ll \omega < E$, the spectrum is dominated by coherent multiple scatterings and scales like $\omega^{-3/2}$.
The power-law fits in each domain are very close to the expected scaling.
In the middle plot, we compare the simulation to the NLL solution with self-consistent $Q_1^2$ from equation \ref{eq:Q1-sf}. 
The ratio between the two is shown in the bottom plot.
There is a good agreement in the LPM region between the simulation and the theory calculation, which we have used as guidance in developing our Monte-Carlo approach.

A comparison of the other channels $g\rightarrow g+g$ and $g\rightarrow q+\bar{q}$ to the theory calculations are shown in Figure \ref{fig:channel_rate}.
For the splitting parton energy much greater than temperature $\omega > 10 T$, the simulation agrees well with the NLL solution. 
Again, please also refer to appendix \ref{app:tune-spectrum} for comparison varying both $E$ and $\alpha_s$ for these two channels.

Finally, we compare the running coupling calculation with the theory curve in Fig. \ref{fig:running} for the $g\rightarrow g+g$ channel.
The theory curves (black lines) are obtained combining equation \ref{eq:AMY-LL} and equation \ref{eq:q3running}.
Different line styles correspond to the variation of the $Q_0$ value around an initial guess $m_D (E/T \ln(E/T) )^{1/4}$ by a factor of $2$ above and below.
For this 1 TeV parton, the scale $Q_0$ is actually very large and the running of $\alpha_s$ is rather slow, which explains the theory curve is not very sensitive to a factor of $4$ change in $Q_0$.
The simulation was performed using the running coupling prescription described in Section \ref{section:running}.
The overall shape of the spectrum in the deep LPM region is again well described by the modified Boltzmann simulation.

\section{Towards phenomenological applications}\label{section:more}
In the previous section, we showed that the modified Boltzmann equation simulation has a good agreement with the theoretical calculation for parton splitting in the LPM regime in an infinite medium.
Towards future phenomenological application, we would like to investigate a few more complex scenarios involving a finite and expanding medium.

\subsection{A finite medium}
For calculation in a finite medium, there is an intricate interference pattern near the boundary which requires solving the original equation using a finite medium temperature profile. 
Or for the case of a thin medium, such effect can be analyzed order by order in the ``opacity ($L/\lambda$) expansion" \cite{Wiedemann:2000za,Gyulassy:1999zd}. 
One important effect in a finite medium is the path-length dependent radiation rate for $L \lesssim \tau_f$. Considering the formation time of very energetic splitting can be comparable to the size of the QGP fireball, the finite size effect is important for heavy-ion collision phenomenology,
Though the modified Boltzmann approach has been constructed to mimic the case in an infinite medium, it does predict an $L$-dependent results in a finite medium and we would like to check whether it is significantly different from the theory expectation.
The origination of the path-length dependence in the modified Boltzmann approach is that the gluons sampled at $t=t_0$ are not considered as independent objects until $t = t_0+\tau_f$, meaning the splitting at time $t$ is initiated by an inelastic vertex at $t-\tau_f$.
In a semi-infinite medium with a step function like temperature profile, 
\begin{eqnarray}
T = \begin{cases}
0 , z<0\\
T_0, z>0
\end{cases}
\end{eqnarray}
there are no scattering centers at $L-\tau_f<0$ and thus introduces a reduction in the radiation rate for small path-length.

\begin{figure}
\includegraphics[width=\columnwidth]{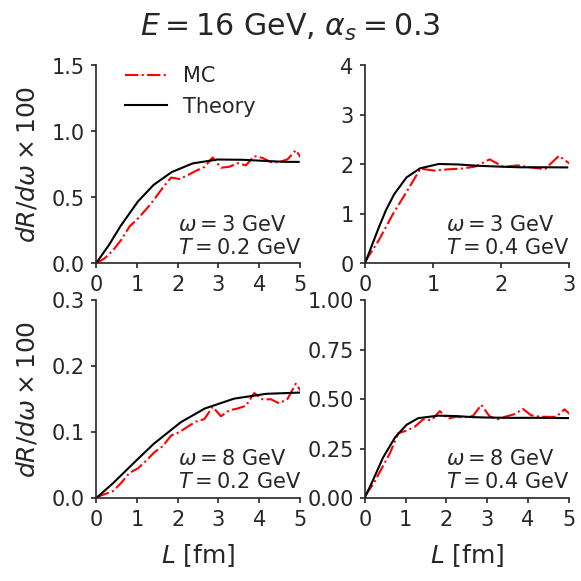}
\caption{Comparison of the path-length dependent rate $dR/d\omega$ from the simulation using $\alpha_s = 0.3$ to the theoretical calculation for splitting $q\rightarrow q+g$ \cite{CaronHuot:2010bp}. The quark energy is $16$ GeV.}
\label{fig:spectra-L-alphas=0.3}
\end{figure}

In Figure \ref{fig:spectra-L-alphas=0.3}, the $q\rightarrow q+g$ rate simulated in a semi-infinite medium is compared to the full calculations obtained in \cite{CaronHuot:2010bp}.
The energy of the quark is 16 GeV, and $\alpha_s = 0.3$.
The medium temperature of the left and the right columns are 0.2 GeV and 0.4 GeV respectively.
Top and bottom rows show the differential rates for the emitted gluon energy $\omega = 3$ GeV and $\omega = 8$ GeV \footnote{In a practical simulation, the rates are obtained by counting gluons within a finite energy range $\omega\pm 0.5$ GeV}.
$dR/d\omega$ is plotted as a function of the path length $L$.
It is evident that theoretical rates (black solid lines) first grow approximately linearly with $L$ and then bend over to transit to $L$-independent ones.
The simulation describes the large $L$ limit well and also approximately captures the point at which the transition happens.
But there are systematic deviations compared to the theory at small path-length.
Therefore it will be of great interest to improve to the current simulation approach at small path-length in the future. 
For example, one possible solution would be using the results from the opacity expansion in the simulation for those splittings that happened close to the boundary and developing matching conditions to the approach we used for the deep LPM regime.

\subsection{An expanding medium}
\begin{figure}
\includegraphics[width=\columnwidth]{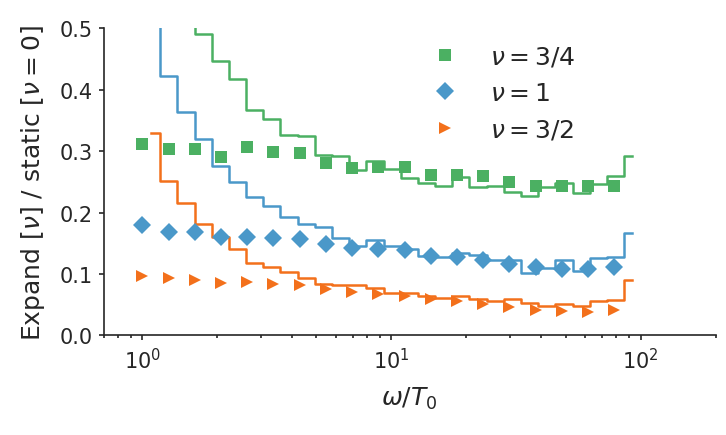}
\caption{The ratios of induced splitting rate in expanding medium to that of a stat medium, with expansion parameter $\nu = 3/4, 1$, and $3/2$. The analytic results are shown in solid lines and simulations denoted as symbols. The conpling constant $\alpha_s=0.3$, the expansion starts at $\tau_0 = 0.2$ fm/$c$ with an initial temperature $T_0 = 1$ GeV.}
\label{fig:Bjorken-BDMPS}
\end{figure}

At the beginning of section \ref{section:modified-Boltzmann}, we have mentioned that the semi-classical transport equation has to be improved when the formation time is comparable to the inverse-gradient of the system.
This is indeed an issue for describing hard parton propagation in realistic medium created in heavy-ion collisions.
Here, we test the modified transport approach in a simple case that only the temporal change of the medium temperature is considered. 
It introduces the inverse medium expansion rate as another time scale $\tau_{\textrm{ex}}$. 
If $\tau_{\textrm{ex}} \lesssim \tau_f$, the multiple-collision effect comes from collision centers along the hard parton trajectory in the time-dependent medium, as is done in the modified transport approach.
In this work, we shall define $\tau_{\textrm{ex}}$ as
\begin{eqnarray}
\tau_{\textrm{ex}} = \left(\frac{d\ln(T^3)}{d \tau} \right)^{-1},
\end{eqnarray}
understood as the time over which the local $\hat{q}\propto T^3$ changes notably.
For simplicity, parametrize the temperature profile as a power law function of the proper time,
\begin{eqnarray}
T(\tau; \nu)^3 = T_0^3\left(\frac{\tau_0}{\tau}\right)^{2-1/\nu},
\label{eq:temperature}
\end{eqnarray}
which mimic the fast-dropping medium temperature at mid-rapidity.
The static case is recovered when $\nu=1/2$ and $\nu=1$ corresponds to the temperature profile of a Bjorken flow \cite{PhysRevD.27.140}.
The resultant $\tau_{\textrm{ex}}$ is
\begin{eqnarray}
\tau_{\textrm{ex}} = \frac{\tau}{2-1/\nu}.
\end{eqnarray}

To compare the response of the modified Boltzmann approach to an expanding medium to theoretical calculations, we use the results obtained in the BDMPS framework \cite{Baier:1998yf} using the power-law type temperature profile in equation \ref{eq:temperature}. 
The splitting probability is,
\begin{eqnarray}
\frac{dP}{d\omega} &=& \frac{\alpha_s}{2\pi E}P_{q\rightarrow qg}(x)\mathfrak{Re}\int_{\tau_0}^{\tau_0+L}\frac{dt_f}{t_f}\int_{\tau_0}^{t_f}\frac{dt_i}{t_i} \frac{1}{\nu^2}\\
\nonumber
&& \left.\left[ I_{\nu-1}(z_i)K_{\nu-1}(z_f)-I_{\nu-1}(z_f)K_{\nu-1}(z_i)\right]^{-2}\right|_{\omega}^{\omega=\infty},\\
z_{i,f} &=& 2i\nu \sqrt{\frac{\hat{q}_g(1-x+C_F/C_A x^2)}{2(1-x)\omega}} \tau_0 \left( \frac{t_{i,f}}{\tau_0}\right) ^{1/2\nu}
\end{eqnarray}
for the $q\rightarrow q+g$ splitting.
For $\nu=1/2$, this expression reduces to the static BDMPS result \cite{Baier:1996kr}. 
As a remark, the BDMPS calculation considers the multiple-soft limit of the collision kernel and therefore does not include the logarithm that comes from the perturbative tail $1/\mathbf{q}^4$. 
Accordingly, we turn off the large-$Q$ matrix-element scatterings and only retain diffusion plus diffusion-induced radiation components in our simulation.
Also, $b=0.75$ is used without the logarithmic correction factor in equation \ref{eq:NLL-b}, and the same $\hat{q}_g = m_D^2 C_A\alpha_s T$ are input to the theory and the simulation.
To suppress other difference in the simulation and the theory, instead of making a direct comparison of the spectrum $dP/d\omega$ to the BDMPS result, we compare the ratio of the splitting probability in an expanding medium to that of a static medium
\begin{eqnarray}
\frac{dP(T=T(\tau;\nu))/d\omega}{dP(T=T_0)/d\omega}
\end{eqnarray}
between simulation and theory to focus on the response to an expanding medium compared to the static case.

The medium expansion starts at $\tau_0=0.2$ fm/$c$ with $T_0=1$ GeV and stops at $\tau = 20$ fm/$c$.
We take four choices of the expansion rate $\nu = 1/2, 3/4, 1, 3/2$, corresponding to a static medium, a slowly expanding medium, Bjorken flow, and a faster-than-Bjorken expansion respectively.
The ratio $R_\nu$ from both theory and simulation are shown in Fig. \ref{fig:Bjorken-BDMPS} for a 100 GeV quark with $\alpha_s=0.3$.
Again, for $\omega/T \gg 1$, the simulation displays the expected decreasing of medium-induced radiation due to the dropping of temperature.
In the future, we are looking forward to making a direct comparison to the solution of equation \ref{eq:full-theory} with both varying temperature and adding medium flow effects.

\section{Comparison with two other Monte-Carlo methods for medium-induced splittings}\label{section:compare}
Before we conclude this work, it is beneficial to compare the current implementation of the medium-induced splittings with two other Monte-Carlo approaches.
We will also summarize the features and caveats of these two methods for readers reference on this subject. 
Also, because the other two methods predict very different radiation spectrum, it is not so instructive to compare their spectrum directly.
Instead, we compare among these approaches the ``energy loss" of a testing quark from gluon radiation,
\begin{eqnarray}
\frac{dE}{dL} = \int_{m_D}^E \omega_g \frac{dR^q_{qg}}{d\omega_g} d\omega_g.
\label{eq:eloss}
\end{eqnarray}
As a remark, this definition is not related to the actually parton / jet energy loss in a medium, but only as a simply way to quantify the difference between different methods.
We shall see that due to different implementations of the medium-induced radiation, the amount of ``energy loss'' is very different between these approaches with the same $\alpha_s$.
As a result, an extraction of interesting medium properties from experimental data using these models can be biased by the way they treat radiative processes.
Therefore, being able to calibrate a model to theoretical calculations as demonstrated in this work is an essential step prior to the comparison to experimental data.

\subsection{The approached used in the improved Langevin equation}
This approach is implemented in the improved Langevin equation \cite{Cao:2013ita}, using a higher-twist calculation of medium-induced single-gluon emission rate and a prescription for multiple emission in a time-evolution manner.
The higher-twist formula is developed in \cite{PhysRevLett.85.3591,Majumder:2009ge} and the single-gluon radiation rate is,
\begin{eqnarray}
\frac{dN_g}{dx d\mathbf{k}^2 dt} = \frac{\alpha_s P(x)\hat{q}_g}{\pi \mathbf{k}^4} 2\left(1-\cos\frac{t-t_0}{\tau_f}\right)
\end{eqnarray}
The rate is time-dependent, coming from the interference between the production of the hard parton at time $t_0$ and one interaction with the medium at time $t$.
From the second emission, a multiple-radiation is implemented by setting the time $t_0$ to be the time of the previous emission so that the probability of the next emission starts to accumulate from zero again as time increases \cite{Cao:2013ita}.
Though it is not immediately clear what this prescription predicts without a simulation, it is possible to get a qualitative understanding by realizing that the typical time separation between two emissions of this model is a time scale $\Delta t = t-t_0$ within which the emission probability is of order one,
\begin{eqnarray}
1 \sim \int_{t_0}^{t} dt\int_{x_c}^1 dx \int d\mathbf{k}^2 \frac{dN_g(t-t_0)}{dx d\mathbf{k}^2 dt}.
\end{eqnarray}
In the soft limit where $P(x) \sim 2/x$, $\tau_f\sim 2xE/\mathbf{k}^2$, and perform the time integral first, then the $\mathbf{k}$ integral with limits from $0$ to $xE$.
The $x$ (with a change of variable to $u = xE\Delta t/2$) integral here divergent at 0.
Though this is not a problem for calculating energy loss if one also included the absorption processes are required by detailed balance.
But to apply the collision rate formalism, one has to render the integral finite with a minimum cut-off $x_c$,
\begin{eqnarray}
1 &\sim& 4\alpha_s\hat{q}\Delta t \int_{x_c}^1 \frac{dx}{x} \int \frac{d\mathbf{k}^2}{\mathbf{k}^4}\left(1-\frac{\sin(\Delta t/\tau_f)}{\Delta t/\tau_f}\right)\\
&=& \frac{\alpha_s\hat{q}_g \Delta t^3}{3u^3} \left( u^3\mathrm{Ci}(u)-3u^2\mathrm{Si}(u) - u^2 \sin(u) \right. \\\nonumber
&&\left. +3u-\sin(u) - 2u\cos(u) \right) \left.\right|_{\frac{\Delta t E x_c}{2}}^{\frac{\Delta t E}{2}} 
\end{eqnarray}
The result can be expanded at small $u$: $\frac{1}{3}\ln(u)-0.752$ and it quickly decays to $0$ at infinity, so a good proxy is to use the small-$u$ expansion but cut-off the upper bound at its zero,
\begin{eqnarray}
1 &\sim&  \frac{\alpha_s\hat{q}\Delta t^3}{3}\ln\frac{2}{ x_c E \Delta t } \propto (g^2 T \Delta t)^3 \ln\frac{2}{ x_c E \Delta t }.
\label{eq:delta-t-formula}
\end{eqnarray}
One sees that typical time between two emissions in this approach is on the order of $1/g^2T$, which is the same as $\lambda$.
Putting typical $\Delta t \sim \lambda$ back to the factor $2(1-\cos(\Delta t/\tau_f))$, the radiation spectrum is indeed suppressed when $\tau_f \gg \lambda$.
However, this suppression is introduced by controlling the correlation between two subsequent emissions, while the LPM suppression actually happens on the level of single emission rate.
Moreover, the logarithmic factor in equation \ref{eq:delta-t-formula} depends on the infrared cut-off \footnote{This cut-off is chosen to be $\omega > \pi T$ in \cite{Cao:2013ita}}, therefore the prediction is cut-off dependent, though logarithmically slow.
This is because that the prescription changes the second emission rate in the same way no matter how soft the previous emission is.
This is, in fact, a feature we have avoided in this work.

\subsection{The ``blocking radiation'' approach}
Another method \cite{ColemanSmith:2012vr} will be termed as the ``blocking radiation approach''.
Similar to this work, the splitting is also first generated through an incoherent process at time $t'$. 
It is then followed by a self-consistent determination of formation time with elastic broadening similar to the procedures described in section \ref{section:modified-Boltzmann}.
However, different from this work, the ``blocking radiation approach'' implements the LPM suppression by requiring that no additional radiation is allowed during the formation time of the previous one, while remembering that in our approach, the emitter can have an arbitrary number of independent ``pre-formed'' final-state copies.
Again, this ``blocking radiation'' approach introduces correlations between subsequent emissions.
A closer investigation reveals a bigger problem.
In an infinite medium, this approach reduces every $\tau_f/\lambda_{inel}$ incoherent emission to one. 
$\lambda_{inel}$ is the mean-free-path of incoherent inelastic collisions, and is related to the elastic mean-free-path by $\lambda_{inel}\propto \lambda_{el}/\alpha_s$. 
It only results in an overall reduction in the radiation spectrum without changing its shape, and the suppression factor $\lambda_{inel}/\tau_f$ is off by a power of $\alpha_s$ compared to the expected one $\lambda_{el}/\tau_f$.

\begin{figure*}
\centering
\includegraphics[width=1.\textwidth]{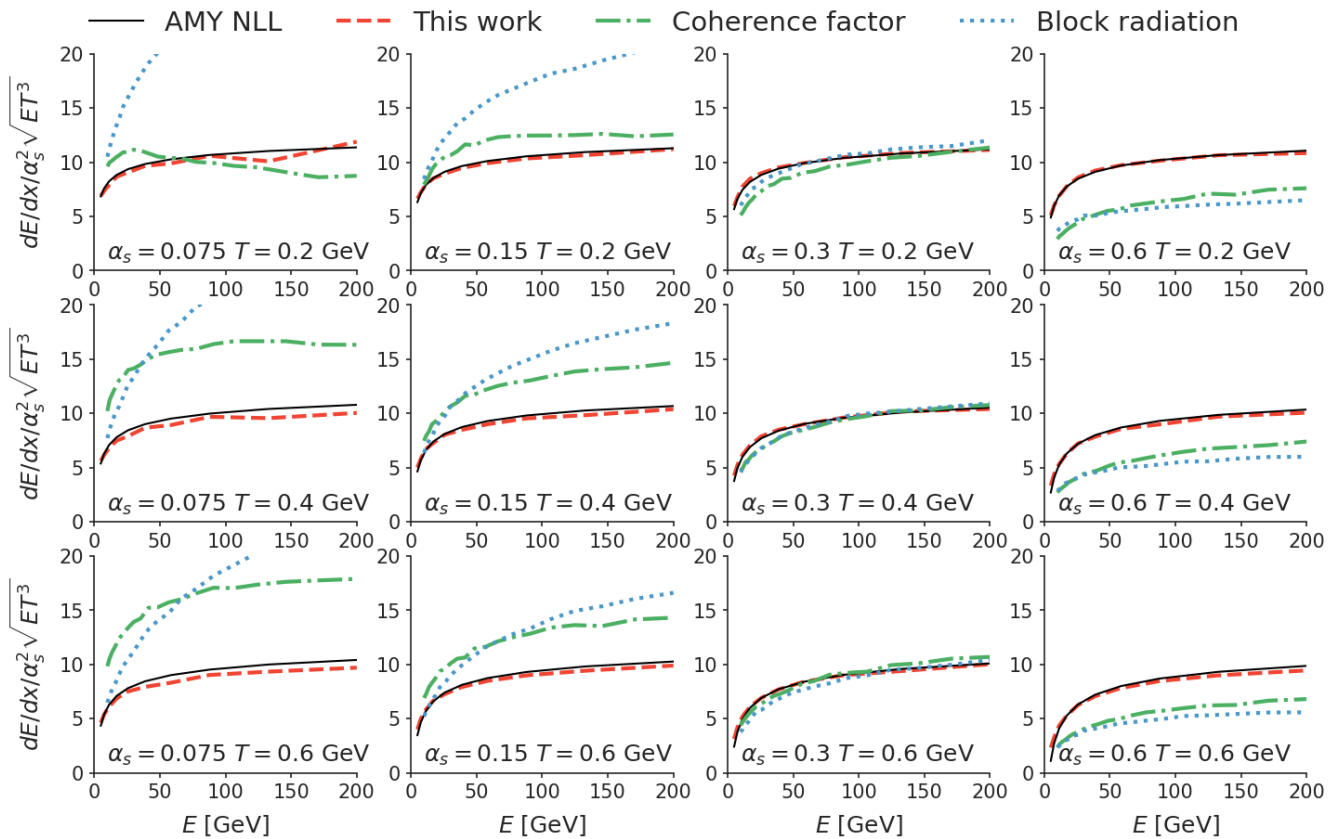}
\caption{Energy loss per unit path lengh $dE/dL$ as a function of energy $E$, temperature $T$ and coupling constant $\alpha_s$. Each column corresponds to a value of the coupling constant $\alpha_s = 0.075, 0.15, 0.3$, and $0.6$ (from left to right). Each row corresponds to a temperature of $T = 0.2, 0.4$, and $0.6$ GeV (from top to bottom). $dE/dx$ is divided by the expected scaling $\alpha_s^2 \sqrt{ET^3}$. The MC implementations of the LPM effect referred to as ``modified Boltzmann", ``coherence factor", and ``blocking radiation" approaches are shown with red-dashed lines, blue-dash-dotted lines, and green-dotted lines respectively. The AMY NLL results are denoted as black boxes.}
\label{fig:eloss-inf}
\end{figure*}

\subsection{Comparison with the modified Boltzmann approach and the analytic results}
In Figure \ref{fig:eloss-inf}, we show the calculation of ``energy loss'' defined in \ref{eq:eloss} of a quark in an ``infinitely large" medium. 
The results presented are normalized by $1/(\alpha_s^2 \sqrt{ET^3})$ in anticipation of the scaling $dE/dL \propto \alpha_s^2 \sqrt{ET^3}$.
For each column, we double the value of $\alpha_s$ and for each row, the temperature is increased by $0.2$ GeV. 
Within each subplot, the parton energy varies from $10$ GeV to $200$ GeV.
The three Monte-Carlo methods of medium-induced energy loss are shown in colored lines and NLL AMY results are shown as black lines.
As expected, the modified Boltzmann approach (red-dashed lines) which describes the radiation spectrum also reproduces the energy, temperature, and coupling constant dependence of the energy loss.
The method used in the improved-Langevin equation (blue-dash-dotted lines) has a similar energy and temperature dependence as the theoretical baseline; however, it systematically deviates from the baseline for different values of the coupling constant in a logarithmic manner.
For the ``block radiation" approaches, the deviations from the baseline regarding their $\alpha_s$-dependence is completely off, which is not surprising as we have discussed its shortcomings.

\section{Summary and outlook}\label{section:summary}
We have investigated the modification to the semi-classical Boltzmann transport equation to include the LPM effect for parton splitting processes in the deep-LPM region, with the guidance from the leading-log and the next-to-leading-log solutions of the AMY equation.
The running coupling effect has also been implemented.
The overall level of agreement between the simulated results and theoretical calculations in the infinite medium limit is promising given the simplicity of this Monte-Carlo procedure. 
Although it is developed for the deep-LPM regime, this approach captures qualitative features of the path-length dependence of medium-induced splittings and the qualitative change of the spectrum shape in an expanding medium compared to the static case.
Future study will focus on improved treatment in a thin medium, and consistent inclusion of the heavy-quark mass effect into the current approach.

Being able to calibrate a Monte-Carlo transport model to the theoretical calculation is important.
As we have demonstrated in section \ref{section:compare}, different modeling of the medium-induced radiation can bias the extraction of the interaction strength between the probe and the medium.
The design of this modified Boltzmann transport approach and its systematic comparison to theoretical calculations allow us to reduce and estimate the uncertainty in these implementations.
This is instrumental for performing an examination of theoretical assumptions and a more meaningful phenomenological extraction of jet transport properties from future model-to-data comparisons in transport model-based studies.

\begin{acknowledgments}
SAB, WK, and YX are supported by the U.S. Department of Energy Grant no. DE-FG02-05ER41367. WK is also supported by NSF grant OAC-1550225.
WK would like to thank Florian Senzel, Jean-Francois Paquet and Yacine Mehtar-Tani for helpful discussions.
\end{acknowledgments}

\begin{appendices}
\section{Approximation of the $2\rightarrow 3$ matrix-elements}
\label{app:23}
Regarding the large-$q$ $2\rightarrow 3$ matrix-elements, in previous study \cite{Ke:2018tsh}, we emploied an improved version of the original Gunion-Bertsch cross-section that works under the limits $\mathbf{k}, \mathbf{q} \ll \sqrt{s}$ and $x \mathbf{q} \ll \mathbf{k}$ \cite{PhysRevD.25.746,Fochler:2013epa,Uphoff:2014hza}.
The original Gunion-Bertsch cross-section \cite{PhysRevD.25.746} only works for soft emissions $x=k^+/p^+ \ll 1$. 
With the improvements made in \cite{Fochler:2013epa,Uphoff:2014hza}, the agreement with the exact matrix-elements is extended to larger $x$, but still the splitting function is only reproduced to $O(x)$.
In the present study, we relax the condition $x \mathbf{q} \ll \mathbf{k}$ in the derivation, so that the full leading order vacuum splitting function can be recovered.
We summarize the matrix-elements here,
\begin{eqnarray}
\overline{|M^2|}_{g+i\rightarrow g+g+i} &=& \overline{|M^2|}_{g+i\rightarrow g+i} P_{gg}^{g(0)}  D_{gg}^{g},\\
\overline{|M^2|}_{g+i\rightarrow q+\bar{q}+i} &=& \frac{C_F d_F}{C_A d_A}\overline{|M^2|}_{g+i\rightarrow g+i} P_{q\bar{q}}^{g(0)} D_{q\bar{q}}^{g},\\
\overline{|M^2|}_{q+i\rightarrow q+g+i} &=& \overline{|M^2|}_{q+i\rightarrow q+i} P_{qg}^{q(0)} D_{qg}^{q},
\end{eqnarray}
where $\overline{|M^2|}_{g+i\rightarrow g+i}$, $\overline{|M^2|}_{g+i\rightarrow g+i}$ and $\overline{|M^2|}_{q+i\rightarrow q+g+i}$ are the spin-color averaged two-body collision matrix-elements with only $\hat{t}$-channel contribution.
Index $i$ represent a medium quark / anti-quark or a medium luon.
The $P_{bc}^{a(0)}(x)$ terms are vacuum splitting functions of parton $a$ to partons $b$ and $c$. 
\begin{eqnarray}
P_{gg}^{g(0)}  &=& g^2  C_A\frac{1+x^4+(1-x)^4}{x(1-x)},\\
P_{qg}^{q(0)} &=& g^2  C_F\frac{1+(1-x)^4}{x},\\
P_{q\bar{q}}^{g(0)} &=& g^2  \frac{N_f}{2}\left(x^2+(1-x)^4\right).
\end{eqnarray}
Finally, the $D_{bc}^{a}$ terms are,
\begin{eqnarray}
D_{qq}^{g} &=& 
C_A(\mathbf{a}-\mathbf{b})^2 + C_A(\mathbf{a}-\mathbf{b})^2 \\\nonumber
&-& C_A (\mathbf{a}-\mathbf{b})\cdot (\mathbf{a}-\mathbf{c}),
\\
D_{q\bar{q}}^{g} &=& 
C_F(\mathbf{a}-\mathbf{b})^2 + C_F(\mathbf{a}-\mathbf{b})^2 \\\nonumber
&-& (2C_F-C_A) (\mathbf{a}-\mathbf{b})\cdot (\mathbf{a}-\mathbf{b}),
\\
D_{qg}^{q} &=& 
C_F(\mathbf{c}-\mathbf{a})^2 + C_F(\mathbf{c}-\mathbf{b})^2 \\\nonumber
&-& (2C_F-C_A) (\mathbf{c}-\mathbf{a})\cdot (\mathbf{c}-\mathbf{b}),
\end{eqnarray}
with the vectors given by
\begin{eqnarray}
\mathbf{a} &=& \frac{\mathbf{k} - x\mathbf{q}}{(\mathbf{k} - x\mathbf{q})^2},\\
\mathbf{b} &=& \frac{\mathbf{k} - \mathbf{q}}{(\mathbf{k} - \mathbf{q})^2},\\
\mathbf{c} &=&  \frac{\mathbf{k}}{\mathbf{k}^2}.
\end{eqnarray}

\section{Energy and coupling constant dependence of the splitting rate}
\label{app:tune-spectrum}
\begin{figure}[t]
\includegraphics[width=\columnwidth]{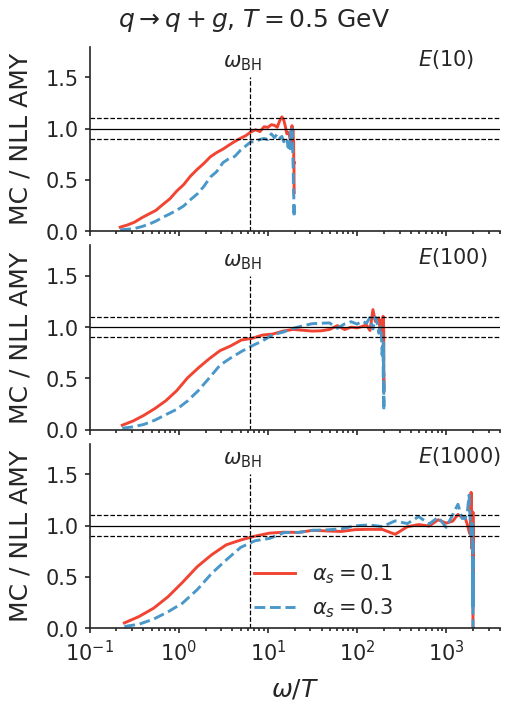}
\caption{Ratios of splitting rate $dR/\omega$ between the modified Boltzmann simulation and the NLL solution for $q\rightarrow q+g$ splitting. The quark energies are $E$ is 10, 100, and 100 GeV from top to the bottom plot. 
And two coupling constants are used: $\alpha_s = 0.1$ (red solid lines) and $\alpha_s = 0.3$ (blue dashed lines).
$\omega$ stands for the gluon energy.
The horizontal dashed lines denote $\pm 10\%$ deviation from unity. }
\label{fig:q2qg}
\end{figure}

\begin{figure}[t]
\includegraphics[width=\columnwidth]{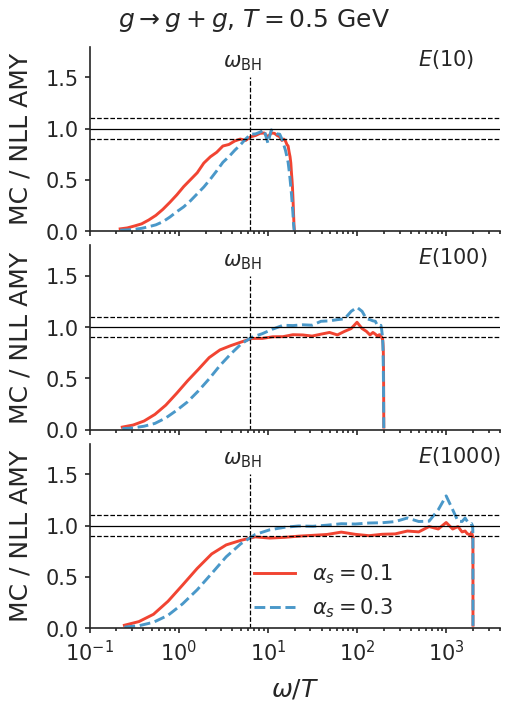}
\caption{The same as Fig. \ref{fig:g2gg}, but for the $g\rightarrow g+g$ splitting, and $\omega$ stands for either energy of the final state gluon.}
\label{fig:g2gg}
\end{figure}

\begin{figure}[t]
\includegraphics[width=\columnwidth]{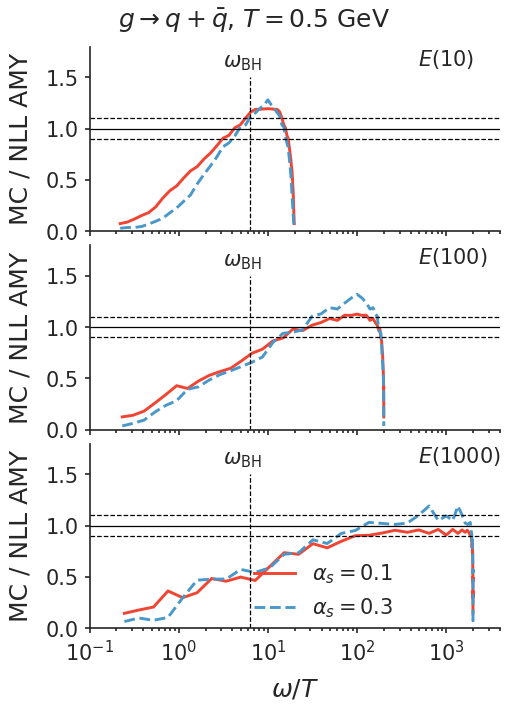}
\caption{The same as Fig. \ref{fig:g2gg}, but for the $g\rightarrow q+\bar{q}$ splitting, and $\omega$ stands the energy of the quark.}
\label{fig:g2qqbar}
\end{figure}

In this appendix, we provide comparisons of splitting rate at different values of energy and coupling constant for the reader's references.
Fig. \ref{fig:q2qg}, Fig. \ref{fig:g2gg} and Fig. \ref{fig:g2qqbar} shows the comparison for channels $q\rightarrow q+g$, $g\rightarrow g+g$ and $g\rightarrow q+\bar{q}$ respectively.
The results are shown as the ratio between the simulations and the NLL solution.
Within each figure, the mother parton energy is 10 GeV, 100 GeV, and 1000 GeV from the top to bottom plot.
We have used two coupling constants at $\alpha_s = 0.1$ (red solid lines) and $\alpha_s = 0.3$ (blue dashed lines).

\end{appendices}
\bibliography{mclpm} 
\end{document}